# Magnetoresistive response of a high mobility 2DES under electromagnetic wave excitation

**R G Mani,**[*,†] **J H Smet,**[†] **K von Klitzing,**[†] **V Narayanamurti,**[*] **W B Johnson,**[‡] **and V Umansky**[∥]

[*]Harvard University, Gordon McKay Laboratory of Applied Science, 9 Oxford Street, Cambridge, MA 02138, USA

[†]Max-Planck-Institut für Festkörperforschung, Heisenbergstrasse 1, 70569 Stuttgart, Germany

[‡]Laboratory for Physical Science, University of Maryland, College Park, MD 20740, USA

[∥]Braun Center for Submicron Research, Weizmann Institute, Rehovot 76100, Israel

**Abstract.** Oscillations of the resistance observed under electromagnetic wave excitation in the high mobility GaAs/AlGaAs 2DES are examined as a function of the radiation frequency and the power, utilizing an empirical lineshape based on exponentially damped sinusoids. The fit-analysis indicates the resistance oscillation frequency, F, increases with the radiation frequency, $\nu$, at the rate $dF/d\nu$ = 2.37 mTesla/GHz; the damping parameter, $\alpha$, is approximately independent of $\nu$ at constant power; and the amplitude, A, of the oscillations grows slowly with the incident power, at a constant temperature and frequency. The lineshape appears to provide a good description of the data.

## 1. Introduction

The magneto-response of the Two Dimensional Electron System (2DES) under electromagnetic wave excitation has been a topic of investigation because it has served to elucidate the collective modes of the low dimensional electronic system.[1] According to Kohn's theorem, cyclotron resonance is independent of electron-electron interactions in a translationally invariant 2DES and consists of a resonance at $h\nu = \hbar\omega_C$ only.[2] Here, $h\nu$ is the photon energy and $\hbar\omega_C$ is the Landau level spacing. At the same time, an infinite 2DES is expected to exhibit a collective plasmon response in the absence of a magnetic field, i.e., B = 0, following the dispersion relation $\omega_p^2 = ne^2k/2\varepsilon_{eff}\varepsilon_0 m^*$, where $\omega_p$ is the plasmon frequency, n is the electron density, e is the electron charge, k is the plasmon wave vector, m* is the effective mass, and for the GaAs/AlGaAs 2DES, $\varepsilon_{eff} = (\varepsilon_{GaAs} + \varepsilon_{vac})/2$, with $\varepsilon_{GaAs}$ = 12.8, and $\varepsilon_{vac}$ = 1.[1,3,4]

The application of a transverse magnetic field leads to a hybridization of cyclotron resonance with the plasmon, producing the magnetoplasmon which follows $\omega_{mp}^2 = \omega_p^2 + \omega_C^2$.[1,5] And, magnetooptical studies have confirmed these dispersion relations by using grating couplers which helped to couple to particular wave vectors, k. [1,4,5]

In a finite size specimen, the length scale established by the boundary can also set the hybridization of cyclotron resonance with the above mentioned plasmon. Vasiliadou et al. investigated the transport signature of this phenomenon in Hall bars and concluded that the selected k is given by the relation k = $\pi$/W, where W is the width of the device.[3] This suggests that the finite sized specimen exhibits the magnetoplasmon or plasmon shifted cyclotron resonance (h$\nu$ = $\hbar\omega_{mp}$) in place of the bare cyclotron resonance (h$\nu$ = $\hbar\omega_C$). This magnetoplasmon contribution supplements edge magnetoplasmons also observed in bounded specimens.[1,6]

It is known that impurity scattering, surface roughness, a Hall electric field, and a boundary, can suffice to make inapplicable Kohn's theorem.[7,8,9] Indeed, impurity scattering, roughness, and a crossed field configuration can lead to transitions also at h$\nu$ = j$\hbar\omega_C$.[8,9] Recent work seems to suggest that short range potential scattering can change the expectation that a finite sized specimen ought to exhibit a magnetoplasmon resonance (h$\nu$ = $\hbar\omega_{mp}$) in place of the bare cyclotron resonance (h$\nu$ = $\hbar\omega_C$).[10]

Here, we report the results of an ongoing experimental investigation that examines,[11] within this context, the magnetotransport response of a high mobility 2DEG under electromagnetic radiation, where the data exhibit a number of resistance

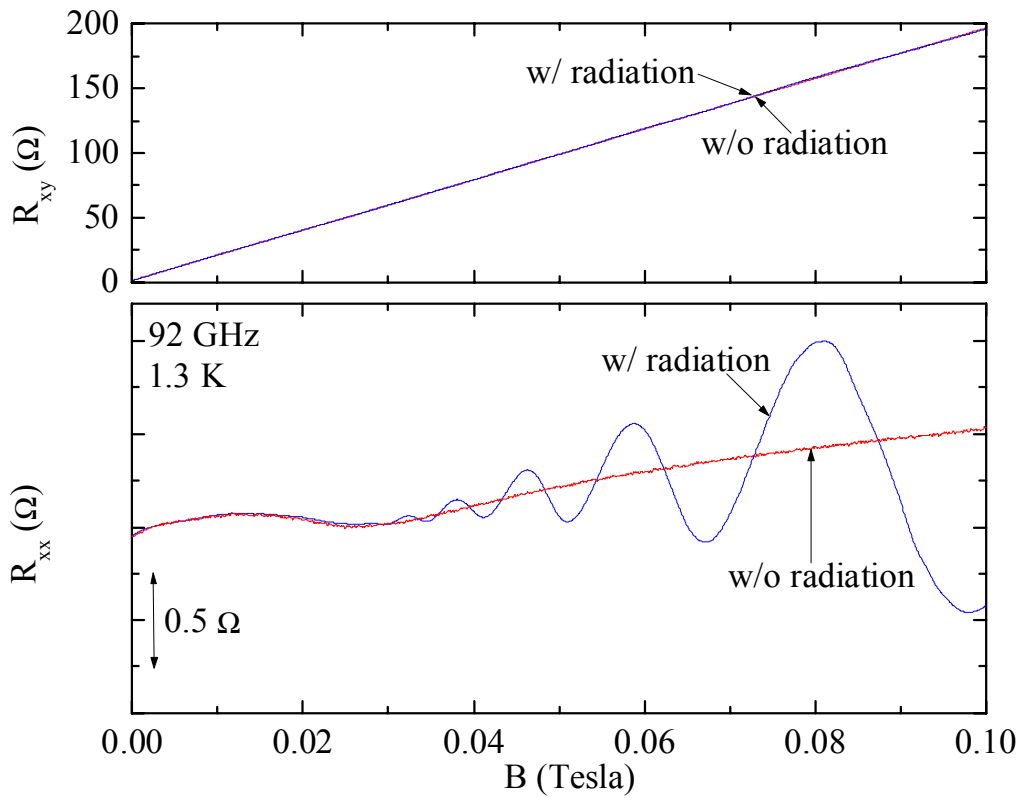

Figure 1) (a) The Hall resistance, $R_{xy}$, as a function of the magnetic field, B, both with (w/) and without (w/o) radiation incident on the sample. (b) The diagonal resistance $R_{xx}$ vs. B with- and without- 92 GHz radiation incident upon the sample. Notably, the resistance under radiation falls below the resistance obtained without radiation in the vicinity of the resistance minima.

oscillations under radiation with periodicity in $B^{-1}$. The observed oscillations appear noteworthy because, at a fixed radiation frequency, many strong resistance oscillations occur below the cyclotron resonance field $B_C$ in high quality specimens, where one expects a reduced density of short ranged scatterers which are typically held accountable for higher order cyclotron resonance type transitions. In this report, we will mainly be concerned with (a) providing an empirical lineshape description for the observed oscillations, (b) extracting electronic characteristics from the fit to the empirical lineshape, and (c) demonstrating that the fundamental oscillation frequency of the waveform coincides with $B_C = 2\pi\nu m^*/e$, i.e., $F = B_C$.

## 2. Experiment

Measurements were carried out on Hall bars and square shaped devices fabricated from low disorder Molecular-Beam-Epitaxy (MBE) grown GaAs/AlGaAs heterostructures. The material was typically characterized by a low density, low mobility condition upon slow cooling in the dark. Brief illumination by a red LED then increased the electron density to $n(4.2\ K) \cong 3 \times 10^{11}\ cm^{-2}$ and the mobility, $\mu$, to $\mu(1.5\ K) \geq 10^7\ cm^2/Vs$. Low frequency lock-in based four-terminal measurements of the resistance and the Hall effect were carried out with the sample mounted inside rectangular waveguides, which served to channel Electro-Magnetic (EM) waves onto the sample, with frequencies in the range $27 \leq \nu \leq 115$ GHz. These EM waves in the microwave part of the spectrum were generated using various sources that covered, piecewise, the above mentioned $\nu$-range, with power levels ranging from a few milliwatts to more than a hundred milliwatts. The EM wave power level was set at the source and subsequently reduced using variable attenuators. Peak power within the waveguide at the sample is thought to be less than a milliwatt in the reported measurements. The sample was immersed in pumped liquid Helium, for temperatures, T, in the range $1.2 \leq T \leq 4.2$ K.

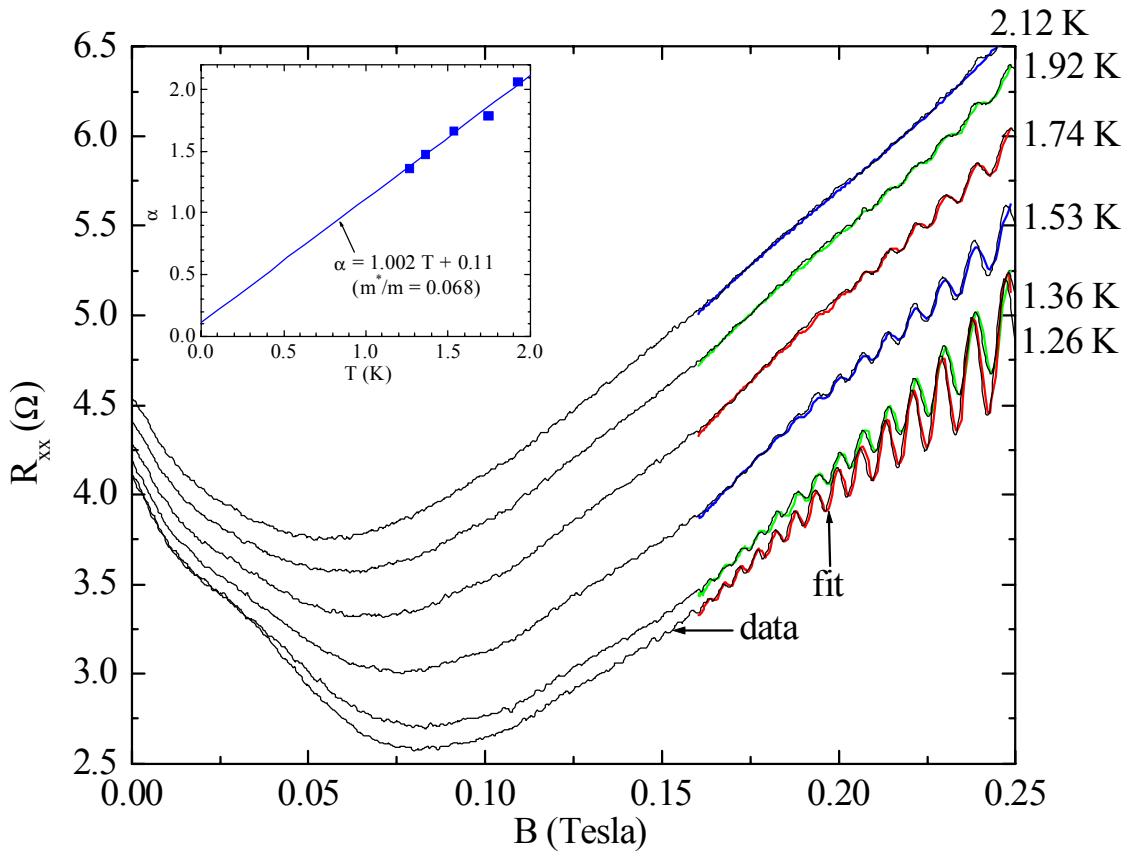

Figure 2) A fit of the Shubnikov - de Haas oscillations in high mobility GaAs/AlGaAs material utilizing the lineshape described in the text. The inset shows a linear variation of the damping parameter, $\alpha$, with the temperature. The electron effective mass determined from the slope of the curve shows good agreement with expectations for the GaAs/AlGaAs system.

## 3. Data and Discussion

Fig. 1 exhibits the transport response of the 2DES over a low magnetic field interval upto 0.10 Tesla both with- and without- radiation at 92 GHz. The figure illustrates that, in these experiments, the application of radiation induces oscillations in the diagonal resistance, $R_{xx}$, approximately symmetrically about the resistance observed in the absence of radiation. Thus, there are B-intervals where the resistance falls below the $R_{xx}$ observed without radiation. At the same time, there does not seem to be much difference between the Hall resistance, $R_{xy}$, obtained with- and without- radiation. Notably, the observed radiation induced oscillatory resistance was found to be insensitive to the magnitude of the current.

Here, we aim to provide an empirical description for the lineshape of these magnetoresistance oscillations and extract information by fitting the data over a finite window. Thus, resistance oscillations will be fit to exponentially damped sinusoids of the type: $A \exp(-\alpha/B) \cos(2\pi F/B + \varphi)$. Here, A stands for the amplitude, $\alpha$ represents the damping parameter, F denotes the frequency, $\varphi$ symbolizes the phase, and B is the magnetic field.

As these high mobility samples exhibited also the Shubnikov-de Haas (SdH) oscillations to relatively low magnetic fields, the data analysis begins by fitting SdH oscillations.[12,13]

The low-T magnetoresistance data *without electromagnetic wave excitation* in a 200 μm wide Hall bar are shown in Fig. 1 to B = 0.25 Tesla. Here, negative magnetoresistance, which increases in magnitude with decreasing temperatures, is observed at the lowest B, along with SdH oscillations at somewhat higher B. Also shown in the figure are lineshape fits to the SdH oscillations based on the semiempirical expression. In the case of SdH oscillations, the temperature variation of the damping parameter $\alpha$ obtained from the fits can be related to effective mass ratio, $m^*/m$, and the Dingle temperature, $T_D$, through the relation $\alpha = \lambda m^*/m(T+T_D)$, where $\lambda$= 14.7 Tesla/K. Hence, the inset depicts the T-dependence of $\alpha$. The observed linear variation indicates that $m^*$ = 0.068m in this specimen, consistent with expectations for GaAs/AlGaAs.[14,15] In addition, the zero-temperature intercept of $\alpha$ yields $T_D \cong 0.1$ K, which provides a measure for the broadening of the Landau levels in this specimen. A comparison of the single particle lifetime $\tau_S = \hbar/2\pi k_B T_D = 1.1 \times 10^{-11}$ s determined from this lineshape analysis of the SdH oscillations with the transport lifetime determined from the dc mobility, $\tau_t = m^*\mu/e = 5.8 \times 10^{-10}$ s, suggests that small angle scattering by long range potentials predominates in these specimens.[7] An inspection of the figure indicates that the empirical lineshape adequately describes the SdH data.

In the following, we apply the same analysis to the radiation induced oscillations, having established the utility of the fit for the SdH effect. In Fig. 3, the magnetoresistance oscillations obtained upon irradiation of the specimen (cf. Fig. 1) with fixed-frequency electromagnetic waves at various power levels have been plotted vs. $B^{-1}$ in order to illustrate periodicity in the inverse magnetic field. Also shown are fits utilizing the empirical lineshape. The figure shows that, while the amplitude of the oscillations increases monotonically with increasing power level, or equivalently, decreasing attenuation, i.e., dB -> 0, the empirical fit continues to provide a good description of the data. The insets in Fig. 3 illustrate the variation of the fit parameters F, the frequency of the oscillations, and A, the amplitude of the oscillations, as a function of the attenuation factor. The left inset shows that the fit-frequency of the oscillations is independent of the power, F (92 GHz) = 0.22 Tesla, which correlates with the constant, power-independent periodicity observed in the data of Figure 3 vs. $B^{-1}$. On the other hand, the amplitude, A, is approximately constant and vanishingly small at the strongest attenuation factors, < -30 dB. It then shows an onset in the vicinity of -25 dB, and subsequently increases approximately linearly as dB → 0. Since the attenuation factor is related to the power level before, $P_{in}$, and after, $P_{out}$, the attenuator by dB = 10 log ($P_{out}/P_{in}$), this indicates that, in the vicinity of, say, -17 dB, an increase in $P_{out}$ by approximately a factor of ten results in just a factor of two change in the amplitude of the resistance oscillations. This suggests that the amplitude of the radiation-induced oscillations is a weak nonlinear function of the incident power.

Figure 4 examines the radiation-induced oscillations as a function of the electromagnetic wave frequency, $\nu$, through lineshape fits using the aforementioned semiempirical waveform. The main panel of figure 4 illustrates that the lineshape continues to provide a good description of the data as the radiation frequency is changed over the interval $100 \leq \nu \leq 107$ GHz. Notably, such behavior also held true all over the entire range $85 \leq \nu \leq 115$ GHz, which constituted the measurement window here. The variation of the interesting fit parameters, F and $\alpha$, is illustrated in the insets of Fig. 4. The left inset shows that the damping factor, $\alpha$, is essentially invariant with the radiation frequency $\nu$, with $\alpha \cong 0.14$ Tesla. This feature can be correlated with the apparent, common exponential envelope for all the oscillatory data in the main panel of Fig. 4. On the other hand, the right inset of Fig. 4 shows that the fit-frequency F increases linearly with the radiation frequency, $\nu$. This feature is a manifestation of the decreasing period with increasing $\nu$ that is observable in the main panel of Fig. 4. Indeed, the rate of change of F with $\nu$, i.e., $dF/d\nu$, can be used to evaluate the effective mass ratio using the relation $dF/d\nu = 2\pi m^*/e$. The observed slope, $dF/d\nu = 2.37$ mTesla/GHz, suggests that $m^*/m = 0.067$, which is also consistent with expectations for the GaAs/AlGaAs system.[14,15] A close inspection of the right inset of Fig. 4, reveals that $F = B_C$, where $B_C = 2\pi\nu m^*/e$, within experimental error.

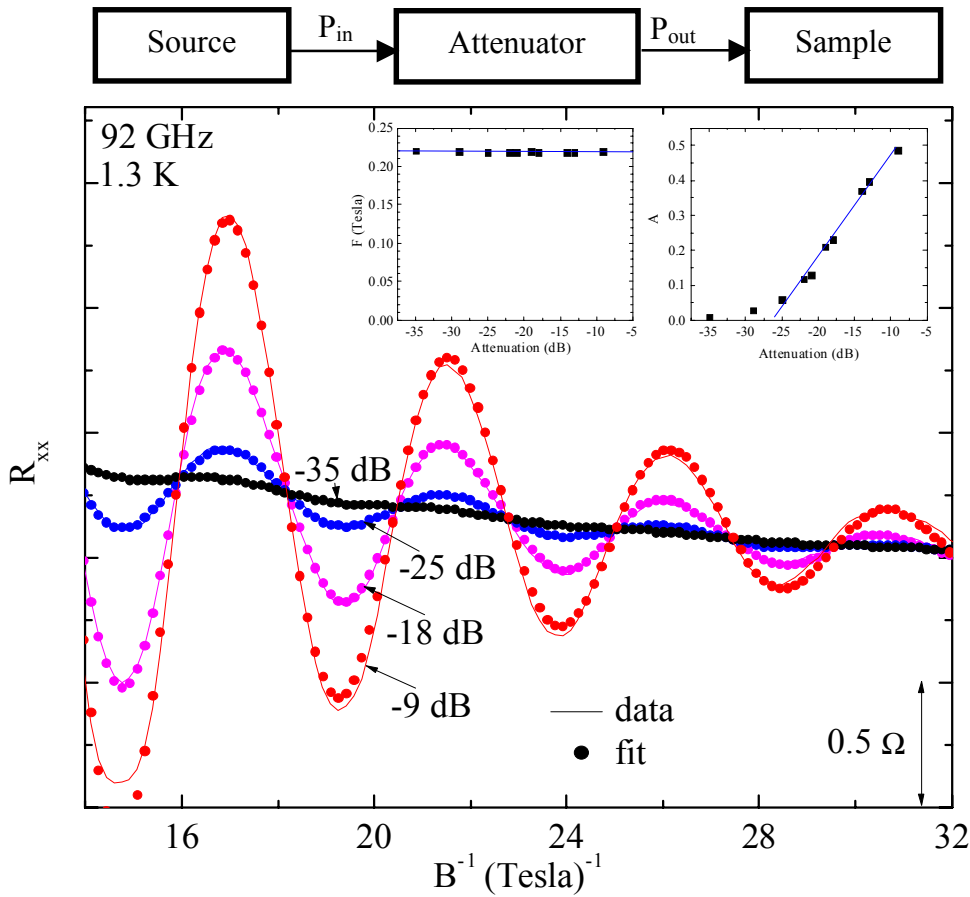

Figure 3) The power dependence of the radiation induced resistance oscillations is illustrated by these data at a few different power attenuation factors. The figure shows the data and the lineshape fit to the empirical expression given in the text. The insets show the variation of the oscillation frequency, F, and the oscillation amplitude, A, vs. the power attenuation factor in dB. A noteworthy feature is that the A increases approximately linearly with the attenuation factor, above a threshold.

The analysis described here shows that resistance oscillations induced by EM

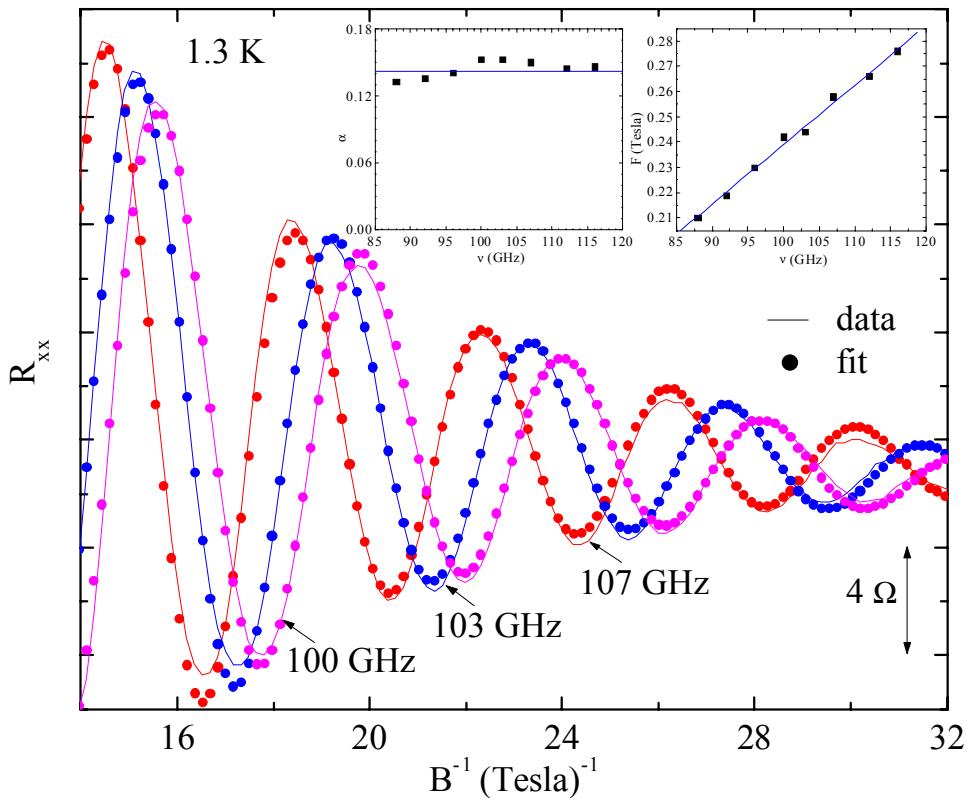

Figure 4) Data at three frequencies $\nu = 100$, 103, and 107 GHz and the corresponding fits of the radiation induced resistance oscillations to the empirical lineshape. The insets show the variation of the fit parameters $\alpha$ and F versus the radiation frequency, $\nu$. The left inset suggests that the damping parameter is independent of $\nu$. This is reflected in the common exponential envelope that is suggested by the data. The right inset shows that the oscillation frequency, F, increases linearly with $\nu$, at the rate $dF/d\nu = 2.37$ mTesla/GHz.

wave excitation in a high mobility 2DES characterized by small angle, long-range-potential scattering can be adequately described by an empirical fit utilizing an exponentially damped sinusoid, as a function of both the radiation power and the radiation frequency, while the change of the resistance oscillation frequency, F, with the microwave frequency, $\nu$, i.e., $dF/d\nu$, reflects the electron effective mass in GaAs/AlGaAs by $dF/d\nu = 2\pi m^*/e$, and the extracted value for $m^*$ shows agreement, within experimental error, with the effective mass determined from a fit of the Shubnikov-de Haas oscillations in the same material. Indeed, the fit frequency F agrees with the magnetic field $B_C$, which suggests that the observed phenomena originates from the interplay between the energy associated with radiation field and the Landau level spacing,[16] although these measurements have been carried out in a low disorder 2DES where the reduced short range scattering would be expected to suppress the occurrence of higher order cyclotron resonance type transitions.[8] It appears that these radiation induced resistance oscillations can be described, in the weak field limit, by $R_{xx}^{osc} = A \exp(-\alpha/B) \sin(2\pi F/B - \pi)$, with $F = B_C$.

The occurrence of resistance oscillations under radiation where the resistance at the minima fall below the resistance in the absence of radiation is noteworthy, and this phenomenon is currently under further investigation. As the Hall angle exceeds $\theta_H > 45^0$ at magnetic fields $B \geq 1$ mTesla in these high mobility specimens, it appears that the $R_{xx}$ reduction at the minima under radiation is due to the suppression of backscattering in the vicinity of the Fermi level. This observation has found recent confirmation in our temperature dependent studies, which suggest that the minima become deeper at lower temperatures, following an activated type law, reminiscent of quantum Hall effect.[17,18]

*R. G. Mani, J. H. Smet, K. von Klitzing, V. Narayanamurti, W. B. Johnson, and V. Umansky, in the Proc. of the 26th International Conference on the Physics of Semiconductors, Edinburgh, Scotland, 29 July - 2 August 2002, IOP Conference Series 171, eds. A. C. Long and J. H. Davies (IOP, Bristol, 2003) H112.*